

\magnification=1200

\font\nrm=cmr10 at 10pt

\font\nnbf=cmbx10 at 12pt
\font\nit=cmti10 at 10pt

\leftskip=4.4truepc
\rightskip=4.2truepc
\baselineskip=22truept

\def\fun#1#2{\lower3.6pt\vbox{\baselineskip0pt\lineskip.9pt
        \ialign{$\mathsurround=0pt#1\hfill##\hfil$\crcr#2\crcr\sim\crcr}}}

{\
\vskip 0.5truecm
\centerline{\nnbf Discrete gauge symmetries in}
\centerline{\nnbf supersymmetric grand unified models}
\vskip 1.truecm
\centerline{M. De Montigny$^{(1,2)}$ $\ $and$\ $ M. Masip$^{(1)}$}
\vskip 0.5truecm
\centerline{\nit $^{(1)}$Institute for Fundamental Theory }
\centerline{\nit Department of Physics, University of Florida}
\centerline{\nit Gainesville, Florida 32611, USA}
\vskip 0.5truecm
\centerline{\nit $^{(2)}$Department of Physics}
\centerline{\nit McGill University, Montreal}
\centerline{\nit Quebec, H3A 2T8, CANADA}
\vskip 1.truecm
\centerline{ABSTRACT}
}

{\nrm
We investigate the presence of discrete gauge symmetries in
Grand Unification models based in $SO(10)$ and $E_6$. These
models include {\it flipped} and {\it unflipped} $SU(5)$,
$SU(4)\!\times\! SU(2)_L\!\times\! SU(2)_R$,
$SU(3)_C\!\times\! SU(3)_L\!\times\! SU(3)_R$,
and $SU(6)\!\times\! SU(2)$. Using the Dynkin formalism we
find the $U(1)$ subalgebras contained in the unified groups, give
an expression for the Higgs fields that preserve each discrete
symmetry, and determine the low-energy matter content implied
by chirality. We discuss two $Z_2$ and three $Z_3$ nonequivalent
cases. Among the possibilities found, only the
usual $Z_2$ matter parity (R-parity) of supersymmetric extensions
is consistent with a minimal matter content with no right- handed
neutrinos, extra Higgs doublets, or nonstandard {\it down}-type quarks.
}
\vfill\eject

\leftskip=1.8pc
\rightskip=1.pc
\baselineskip=21truept

{\leftline {\bf 1. Introduction\/}}
\medskip

In supersymmetric (SUSY) extensions of the standard model (SM)
 with minimal matter content
 it is possible to define dimension-four operators
 that violate lepton and baryon number ($L$ and $B$) [1].
 If present, these terms would
 produce proton decay mediated by SUSY partners of quarks and leptons
 and other unobserved processes.
 To prevent this, one usually assumes the presence of a discrete
 symmetry of the superpotential known as matter parity [1,2].
 The R-parity of the minimal extension (with {\it even} standard
 fields and {\it odd} SUSY partners), in particular,
 corresponds to the $Z_2$ matter parity that changes the sign of
 quark and
 lepton superfields while leaving the two Higgs doublets
 unchanged. It has been recently suggested, however, that other $Z_N$
 symmetries forbidding the same trilinears in the superpotential [2],
 and even discrete symmetries
 allowing $L$ or $B$ violation (baryon and lepton parities [3]),
 may define phenomenologically consistent models.

A possible explanation for the origin of these discrete symmetries
 is provided by Grand Unification Theories (GUTs) [4], where they
 could appear in the following way
 as a remnant of a gauge symmetry.
 Suppose that the
 GUT Lie group contains an extra $U(1)$, with gauge charges $\ Q_i\
 \epsilon\ Z\ $
 and $\ Q_H\ne 0\ $ for the standard superfields $\Phi_i$ and
 GUT Higgs $H$, respectively. If $\ Q_H=0\ mod\ N\ $,
 a vacuum expectation value (VEV) of $H$ will
 break the $U(1)$ factor but will imply an effective model
 still invariant under the gauge transformation
 $\ \Phi_i\longrightarrow exp(i2\pi Q_i/N)\Phi_i\ $
 [note that $\ exp(i2\pi Q_H/N)=1\ $],
 which in general defines a $Z_N$ discrete symmetry.
 Moreover,
 it has been argued that this is the only type of global symmetry
 not anomalous with respect to gravitational (wormholes, etc.) effects [5].

It is well known that the usual R-parity may be obtained from
 models containing a $U(1)_{B-L}$ factor, such as in $SO(10)$.
 The conditions (based on the congruence class
 of the order parameters)
 for this to happen have been recently discussed in Ref.~[6]. In addition,
 Ib\'a\~nez and Ross [3] have classified all the discrete symmetries
 of phenomenological interest in SUSY models, and have established
 consistency conditions that must be satisfied if these
 symmetries are to be a subgroup of a non-anomalous $U(1)$ gauge symmetry.

In this paper we study
 some usual GUT scenarios, discussing all the gauged $Z_N$
 symmetries that may appear in their effective low-energy limit.
 As we mentioned above, they are a consequence of the
 choice of Higgs fields used to break the extra gauge symmetry.
 We shall make extensive use of the Dynkin formalism [7,8]
 to identify and classify the GUT Higgs fields leading
 to each discrete symmetry.
 We also determine, for each case,
 the matter content implied by chirality under
 $SU(3)_C\times SU(2)_L\times U(1)_Y\times Z_N$. Note that
 nonstandard fields that usually appear in GUT models
 in vectorlike representations of the SM gauge symmetry
 (right-handed neutrinos, pairs of {\it down} type quarks, etc.)
 may become here chiral due to the extra $Z_N$ factor.

The next section is devoted to $SO(10)$ and its subgroups
 [{\it flipped} and {\it unflipped} $SU(5)$ and the
 left-right symmetric models $SU(4)\times SU(2)^2$ and
 $SU(3)\times SU(2)^2\times U(1)$]. We then consider in
 Section 3 models based on $E_6$, and make some remarks about
 other extensions.

\medskip
\leftline {\bf 2. Models based on $SO(10)$}
\medskip

 $SO(10)$ is the simplest GUT gauge group
 containing the standard $SU(3)\times SU(2)\times U(1)$ group
 of symmetry
 with {\it (i)} chiral multiplets able to accommodate the
 spectrum of quarks and leptons and {\it (ii)} all its
 representations free of anomalies [4].
 The {\bf 16} irreducible representation (or {\it irrep\/}) contains the
 standard fermions of one family plus a right-handed neutrino,
 while the electroweak Higgs
 fields are usually assigned to the vectorlike irreps
 {\bf 10}, {\bf 120}, and/or {\bf 126}.
 After the
 extra symmetry is broken, this scenario is consistent with
 a minimal matter content of three families of quarks and
 leptons which are light because of chirality plus two
 Higgs doublets protected of heavy mass contributions by
 some other reason (this is the GUT hierarchy problem).

The Cartan subalgebra of $SO(10)$ has dimension five.
 That means that in $SO(10)$ there are a maximum of five
 simultaneously diagonalizable generators.
 In the Dynkin formalism [7,8] these generators label
 each element of an irrep by a set of five integers, or
 {\it weight vector\/}.
 The set of weight vectors in an irrep
 is easily obtained from the highest weight, which specifies the
 representation, by subtracting a finite number of roots.
 The roots (in the basis of fundamental weights) correspond to
 rows of the Cartan matrix, that can in turn be derived from the
 corresponding Dynkin (or {\it Coxeter-Dynkin\/}) diagram. As
 we shall see, this
 formalism is particularly convenient for the analysis of
 $U(1)$s and gauge discrete symmetries.
 Hereafter, we shall follow the notation of Ref.~[8].

The {\bf 16} irrep of $SO(10)$ has highest
 weight vector $\ (0\ 0\ 0\ 0\ 1)\ $; its weight system is listed in
 Table I. We can define five independent charges as
 linear combinations of the basis elements of the Cartan
 subalgebra and embed the SM in $SO(10)$. We consider
 $\ A=[1\ 2\ 2\ 1\ 1]\ $; $\ B={1\over \sqrt{3}}
 [-\! 1\ 0\ 0\ 1\ -\! 1]\ $
 ; $\ C={1\over 2}[0\ 0\ 1\ 1\ 1]\ $;
 $\ D={1\over 3}
 [-\! 2\ 0\ 3\ -\! 1\ 1]\ $; $\ E=[2\ 0\ 2\ 1\ -\! 1]\ $,
 where they are specified in the dual basis
 (the charge $Q$ of a weight
 vector $\lambda$ will be obtained from the scalar product
 $Q\cdot\lambda$).
 The charges $A$ and $B$ are the two
 diagonal generators $\lambda_3$ and $\lambda_8$
 of $SU(3)_C$, while
 $C$ is the standard weak isospin $I^W_3$ of $SU(2)_L$.
 For the weak hypercharge $Y$ one may use $\ Y=D\ $ or
 $\ Y=-{1\over 5}(D+2E)\ $. These two assignations of
 hypercharge are equivalent in the sense that they are related
 by a {\it Weyl reflection\/} [7] of the root system of
 $SU(2)_R\subset SO(10)$,
 but they lead to different scenarios when $SO(10)$ is projected
 down to models with less symmetry. In particular,
 they imply {\it unflipped} and {\it flipped} $SU(5)$ [9],
 respectively.
 We define $\ Y=D\ $ and, when discussing the different
 subgroups of $SO(10)$, we shall include the Weyl reflection in the
 projection matrices (see below). The flavors contained in the
 {\bf 16} and ${\bf 10}=(1\ 0\ 0\ 0\ 0)$ irreps are given
 in Table I and II.

The nonstandard $U(1)$ in $SO(10)$ is generated by
 any combination of charges that contains $E$.
 Following the procedure of Ref.~[3], we use the weak hypercharge to
 make zero the nonstandard charge $Q_1$ of $\ q\equiv (u\ d)\ $.
 Conveniently normalized, $Q_1$ is then given by
 $$Q_1=-{3\over 5} D-{1\over 5} E = [0\ 0\ -\! 1\ 0\ 0]\ .\eqno(1)$$
 The charges of quark, lepton, and Higgs superfields
 in the {\bf 16} and {\bf 10} irreps are
 obtained from the product $Q_1\cdot\lambda$, where
 $\lambda$ is the corresponding weight vector. We obtain \vfill\eject
 $$\eqalign{Q_1(q, u^c, d^c, l, e^c, N)&=(0,1,-1,0,-1,1)\ ,\cr
 Q_1(h, h', D, D^c)&=(-1,1,0,0)\ ,\cr}\eqno(2)$$
 where $l\equiv (\nu\ e)$, $h\equiv (h^+\ h^0)$, and $h'\equiv
 (h'^0\ h'^-)$.
 The notation in Eq.~(2) should not be confused with that
 of a weight vector.
 Higgs doublets in the {\bf 120} or {\bf 126} irreps would
 have the same $Q_1$ charges.

The $U(1)$ symmetry generated by $Q_1$ contains the
 generic $Z_N$ discrete symmetry
 $$\Phi_i\longrightarrow exp\left (i{2\pi\over N}Q_1\right )\Phi_i
 \ ,\eqno(3)$$
 whose action on all the fields in the
 {\bf 16} and {\bf 10} is given in Table III
 (there we also give the particular cases with $N=2$ and $N=3$).
 The $Z_N$ symmetry will survive to low energy if the
 Higgs $H$ that breaks the extra $U(1)$ satisfies
 $$Q_1(H)=0\ \ mod\ N\ .\eqno(4)$$
 Note that one needs at least one Higgs with
 $\ Q_1(H)\ne 0\ $ to break the $U(1)$ and reduce the rank from
 five to four. On the other hand, $H$ must be neutral with respect to the
 SM symmetry; imposing zero $A$, $B$, $C$, and $D$ charges we
 find the expression for the weight vector of a generic
 GUT Higgs in $SO(10)$:
 $$H_n=(-\! n\ n\ -\! n\ 0\ n)\ ,\eqno(5)$$
 where $n$ is an integer.
 The charge $Q_1$ [see Eq.~(1)] of such a field is
 $$Q_1(H_n)=n\ .\eqno(6)$$
 A VEV $<\! H_{1}\! >$ would break all of the possible $Z_N$
 symmetries ($Z_1$ is the identity).
 For $n\! >\! 1$, $<\! H_{n}\! >$ breaks the extra $U(1)$
 of $SO(10)$ while leaving unbroken a $Z_n$ discrete
 symmetry (or, more precisely, a $Z_N$ with $n=0\ mod\ N$).

We find that the smallest ({\it i.e.} lowest dimensional) irrep
 containing $H_n$ has highest weight
 $$\Lambda_n =(0\ 0\ 0\ 0\ n)\ ,\eqno(7)$$
 and dimension
 $$dim\ \Lambda_n =(1+n)(1+{n\over 2})(1+{n\over 3})^2
 (1+{n\over 4})^2(1+{n\over 5})^2(1+{n\over 6})(1+{n\over 7})\ .
 \eqno(8)$$
 Note that each $H_n$ is contained in many different irreps.
 We list in Table IV some irreps
 (their highest weight and the dimension) containing $H_n$ for
 $n=1, 2, 3$. If, for example, one uses the SM singlet
 $H_1=(-\! 1\ 1\ -\! 1\ 0\ 1)$ in the
 ${\bf 16}=(0\ 0\ 0\ 0\ 1)$
 representation to break
 $SO(10)$, no discrete symmetry survives,
 whereas using the adequate weights in the
 ${\bf 126}=(0\ 0\ 0\ 0\ 2)$ and
 $\overline {\bf 672}=(0\ 0\ 0\ 0\ 3)$
 one obtains the
 $Z_2$ and the $Z_3$ symmetries in Table III, respectively.
 Obviously, the GUT Higgs field $H_0=(0\ 0\ 0\ 0\ 0)$,
 like the singlets in the
 ${\bf 45}=(0\ 1\ 0\ 0\ 0)$, ${\bf 54}=(2\ 0\ 0\ 0\ 0)$
 and ${\bf 210}=(0\ 0\ 0\ 1\ 1)$ irreps of $SO(10)$, do not break any
 discrete symmetry. However, their VEVs do not reduce the rank of
 the gauge group either (they would lead to intermediate rank-five
 models).

Each of the $Z_N$ symmetries implies a definite spectrum of light
 fields and a different set of
 couplings in the superpotential.
 In particular, a Majorana mass term for the right handed neutrino
 in the {\bf 16} transforms under $Z_N$ like $\alpha^{-2}$, and
 it is forbidden for $N\ne 2$. Consequently, $SO(10)$ models with an
 unbroken $Z_{N > 2}$ symmetry will include one non-weakly-interacting
 neutrino per family.
 Note that the {\it down} type quarks
 $D$ and $D^c$ (and, in general, all the fields in the multiplet
 of the electroweak Higgs) may receive heavy mass contributions of
 order $<\! H_n\! >$ and should not appear in the effective
 low energy model.

Considering the terms in the superpotential $P$,
 we find that all the $Z_N$ symmetries
 allow the standard Yukawa couplings necessary to give mass to
 quarks and leptons,
 $$P_{MSSM}= y_u\ q u^c h\ +\ y_d\ q d^c h'\ +\
 y_e\ l e^c h'\ +\ \mu\ h h'\ ,\eqno(9)$$
 and forbid the $B$ and $L$ violating terms
 $$P'= y_1\ u^c d^c d^c\ +\ y_2\ q d^c l\ +\
 y_3\ l l e^c\ +\ \mu'\ l h\ .\eqno(10)$$
 The $Z_2$ discrete symmetry in Table III is equivalent to the usual
 matter parity, which would be obtained
 from the product with
 the $Z_2$ symmetry in $U(1)_Y$
 $(q, u^c, d^c, l, e^c, h, h')
 \rightarrow (-q, u^c, d^c, -l, e^c, -h, -h')$.
 The differences between $Z_N$ models will only appear in higher
 dimensional operators (effective nonrenormalizable terms) [2]
 and in the couplings of the extra neutrino $N$. The model with
 $Z_3$ symmetry, for example, will contain in $P$
 $$P_3=y_\nu\ l h N\ +\ \lambda\ N N N\ .\eqno(11)$$
 The couplings $y_\nu$ must be small enough since they give Dirac
 masses to the SM neutrinos. The trilinear $N^3$ violates
 $L$ [$lhN\ $ defines $L(N)=-1$] and $R$-parity, allowing
 the $L$-violating decay of the lightest SUSY particle (LSP) into
 quarks and leptons. Some phenomenological implications of
 models with $Z_3$ matter parity have been explored in Refs.~[2,10].

The group $SO(10)$ includes
 $\ G_1=SU(5)\times U(1)\ $,
 $\ G'_1=SU(5)\subset G_1\ $,
 $\ G^R_1=SU(5)\times U(1)_{flipped}\ $,
 $\ G_2=SU(4)\times SU(2)_L\times SU(2)_R\ $,
 $\ G'_2=SU(3)_C\times SU(2)_L\times SU(2)_R\times U(1)_{B-L}
 \subset G_2\ $.
 Each one of these groups contain the SM symmetry and adequate
 chiral representations, and could define by themselves suitable
 GUT models. However, to understand the cancellation
 of anomalies or the unification of the three gauge couplings, one
 has to embed them in $SO(10)$.

A semisimple subgroup $G_i$ of $SO(10)$ is specified by the
 projection matrix $P_{G_i\subset SO(10)}$ (see details in
 Ref.~[11]), that projects
 weights $\lambda_{SO(10)}$ of $SO(10)$ onto weights $\lambda_{G_i}$
 of $G_i$,
 $$\lambda_{G_i} = P_{G_i\subset SO(10)}\cdot \lambda_{SO(10)}\ .
 \eqno(12)$$
 The dual vectors $Q$ used to specify the charges satisfy
 $$Q_{SO(10)}=Q_{G_i}\cdot P_{G_i\subset SO(10)}\ .\eqno(13)$$

For a sequence of subgroups $G'\subset G\subset SO(10)$, we have
 $P_{G'\subset SO(10)}=P_{G'\subset G}\cdot P_{G\subset SO(10)}$.
 Also, since we have fixed the embedding of the SM in $SO(10)$,
 we shall include in the projection matrices
 the Weyl reflection $R$ mentioned above.
 If $R$ (written as a matrix) cannot be reduced to a Weyl reflection
 of the subgroup
 $G_i$ then $P_{G_i\subset SO(10)}$ and
 $P_{G_i\subset SO(10)}\cdot R\equiv P_{G^R_i\subset SO(10)}$
 will define nonequivalent models.

 For non-semisimple subgroups, it is convenient to give
 the charge of $SO(10)$ that corresponds to the $U(1)$ factor
 and express the charge of the projected weight.
 The projection matrices and $U(1)$ charges of the subgroups under
 consideration \break are [8]
 $$\eqalign{P_{G_1,G'_1\subset SO(10)}&=
 \pmatrix{1&1&0&0&0\cr
 0&0&1&0&1\cr
 0&0&0&1&0\cr
 0&1&1&0&0\cr},\ E=[2\ 0\ 2\ 1\ -\! 1]\ ;\cr
 &\cr
 P_{G^R_1\subset SO(10)}&=
 \pmatrix{1&1&1&0&0\cr
 0&0&0&0&1\cr
 0&0&1&1&0\cr
 0&1&0&0&0\cr},\ E^R=[2\ 0\ -\! 2\ 1\ -\! 1]\ ;\cr
 &\cr
 P_{G_2\subset SO(10)}&=
 \pmatrix{1&1&1&0&1\cr
 0&1&1&1&0\cr
 -\! 1&-\! 1&-\! 1&-\! 1&0\cr
 0&0&1&1&1\cr
 0&0&1&0&0\cr};\cr
 &\cr
 P_{G'_2\subset SO(10)}&=
 \pmatrix{
 1&1&1&0&1\cr
 0&1&1&1&0\cr
 0&0&1&1&1\cr
 0&0&1&0&0\cr},\
 Q_{B-L}={1\over 3}[-\! 2\ 0\ 0\ -\! 1\ 1]\ .\cr
 &\cr
 }\eqno(14)$$
 The projection to {\it flipped} $SU(5)$
 $P_{G^R_1\subset SO(10)}$ has been obtained from
 $$P_{G^R_1\subset SO(10)}=
 P_{G_1\subset SO(10)}\cdot R_3\ ,\eqno(15)$$
 where $R_3$ is the Weyl reflection with respect to the simple
 root $\alpha_3$ of $SO(10)$
 $$R_3=\pmatrix{1&0&0&0&0\cr
 0&1&1&0&0\cr
 0&0&-1&0&0\cr
 0&0&1&1&0\cr
 0&0&1&0&1\cr}\ \eqno(16)$$
 that exchanges $u^c\leftrightarrow d^c$, $e^c
 \leftrightarrow N$, and $(h^+\ h^0)\leftrightarrow
 (h'^0\ h'^-)$.

The subgroups $G_1$,
 $G^R_1$, $G_2$, and $G'_2$ still contain
 the extra $U(1)$ previously considered.
 The Higgs VEVs leaving a $Z_N\!\subset\! U(1)$ unbroken can
 be found by projecting the weights $H_n$ of $SO(10)$ in Eq.~(5).
 For $G_1$, $H_n=(0\ 0\ 0\ 0)$ and $E(H_n)=-5n\ $; for
 $G^R_1$, $H_n=(-\! n\ n\ -\! n\ n)$ and $E^R(H_n)=-\! n\ $;
 for $G_2$, $H_n=(0\ 0\ n)(0)_L (-\!n)_R\ $; and for
 $G'_2$, $H_n=(0\ 0)_C (0)_L (-\!n)_R\ $ and $Q_{B-L}=n$.
 In a {\it flipped} $SU(5)$ ($G^R_1$) model derived from $SO(10)$,
 for example,
 $$\eqalign{
 H_1&= (-\! 1\ 1\ -\! 1\ 1)\in {\bf 10}=(0\ 1\ 0\ 0)\ ,
 \ \ E^R(H_1)=-1\ ;\cr
 H_2&= (-\! 2\ 2\ -\! 2\ 2)\in \overline{\bf 50}
 =(0\ 2\ 0\ 0)\ ,\ \ E^R(H_2)=-2\ .
 \cr}\eqno(17)$$
 Therefore, if the VEVs of the SM singlet in the
 ${\bf 10}(-1)$ representation of {\it flipped} $SU(5)$ are used
 to break the extra symmetry,
 no gauge $Z_N$ based on $SO(10)$ survives, whereas GUT Higgs
 in the $\overline{\bf 50}(-2)$ leaves a $Z_2$ discrete
 symmetry
 (the usual matter parity) unbroken. Analogous conclusions can be
 extracted for the other subgroups.

The rank-four subgroup $SU(5)$ ($G'_1$)
 may result from one of the breakings
 of $SO(10)$ that preserves a $Z_N$;
 in that case, to break $SU(5)$ and still preserve the $Z_N$
 one must avoid certain representations. For instance,
 the SM singlet $(0\ 0\ 0\ 0)$ in the ${\bf 24}=(1\ 0\ 0\ 1)$
 irrep of $SU(5)$ that results from projecting the
 $H_1\in {\bf 144}=(1\ 0\ 0\ 1\ 0)$ of $SO(10)$ (see Table II)
 would break any possible $Z_N$ discrete symmetry.

\medskip
\leftline {\bf 3. Models based on $E_6$}
\medskip

The exceptional group $E_6$ defines another anomaly-free GUT with
 adequate chiral representations. The fundamental irrep
 ${\bf 27} = (1\ 0\ 0\ 0\ 0\ 0)$ (see Table V) contains a
 family of quarks and leptons, the two Higgs doublets ($h$ and
 $h'$) needed in SUSY models, two down-type quarks ($D$ and $D^c$),
 and two non-weakly-interacting neutrinos ($\nu_4$ and $\nu_5$).
 Like in $SO(10)$,
 from a model with three ${\bf 27}$ multiplets
 one may obtain a low-energy limit with
 minimal matter content,
 since all the nonstandard fields are in vectorlike representations
 of the SM symmetry and
 should become massive at the high scales of gauge
 symmetry breaking.

The rank-6 group $E_6$ contains $SO(10)\times U(1)$ as a subgroup;
 the projection matrix and $U(1)$ charge are
 $$P_{SO(10)\subset E_6}=
 \pmatrix{
 0&1&1&1&0&0\cr
 0&0&0&0&0&1\cr
 0&0&1&0&0&0\cr
 0&0&0&1&1&0\cr
 1&1&0&0&0&0\cr},\
 F=[1\ -\! 1\ 0\ 1\ -\! 1\ 0]\ .\eqno(18)$$
 The five charges $A,...,E$ in $E_6$ can be obtained
 from the
 corresponding charges in $SO(10)$ [see Eq.~(13)], giving
 $\ A=[1\ 2\ 3\ 2\ 1\ 2]\ $;
 $\ B={1\over \sqrt{3}}[-\! 1\ -\! 2\ -\! 1\ 0\ 1\ 0]\ $;
 $\ C={1\over 2}[1\ 1\ 1\ 1\ 1\ 0]\ $;
 $\ D={1\over 3}[1\ -\! 1\ 1\ -\! 3\ -\! 1\ 0]\ $;
 $\ E=[-\! 1\ 1\ 4\ 3\ 1\ 0]\ $. We list in Table V all the weights
 in the ${\bf 27}$ irrep which, under $SO(10)\times U(1)$,
 decomposes as ${\bf 27}\rightarrow {\bf 16}(1)+{\bf 10}(-2)+
 {\bf 1}(4)$.

The two nonstandard charges in $E_6$ are, conveniently normalized,
 $$\eqalign{
 Q_1 = & - {3\over 5} D - {1\over 5} E = [0\ 0\ -\! 1\ 0\ 0\ 0]\ ,\cr
 Q_2 = & F + {1\over 5} E - {12\over 5} D =
 [0\ 0\ 0\ 1\ 0\ 0]\cr}\eqno(19)$$
 [$Q_1$ corresponds to the $SO(10)$ charge given in Eq.~(1)].
 For the fields in the ${\bf 27}$ it gives
 $$\eqalign{Q_1(q, u^c, d^c, l, e^c, h, h', D, D^c, \nu_4,
 \nu_5)&=
 (0,1,-1,0,-1,-1,1,0,0,1,0)\ ,\cr
 Q_2(q, u^c, d^c, l, e^c, h, h', D, D^c, \nu_4, \nu_5)&=
 (0,1,0,1,-1,-1,0,0,-1,0,1)\ .\cr}\eqno(20)$$
 The combination of the two $U(1)$s defined by these charges
 contains a generic $Z_N\times Z_M$ discrete symmetry
 given in Table VI in terms of
 $g^{(1)}_N\in Z_N\subset U(1)_{Q_1}$ and
 $g^{(2)}_M\in Z_M\subset U(1)_{Q_2}$, that
 represent the generators of the first and second factors
 of the discrete symmetry group, respectively.
 Thus an element of $Z_N\times Z_M$
 has the generic form
 $(g^{(1)}_N)^n\cdot (g^{(2)}_M)^m$, where $n<N$ and $m<M$.
 Among these symmetries, we identify three different $Z_2$
 [generated by $g^{(1)}_2$, $g^{(2)}_2$
 and $g^{(1)}_2\cdot g^{(2)}_2$]
 and four $Z_3$ [$g^{(1)}_3$, $g^{(2)}_3$, $g^{(1)}_3\cdot g^{(2)}_3$,
 and $g^{(1)}_3\cdot (g^{(2)}_3)^{-1}$].
 Some of them, however, are
 related by the Weyl reflection $R=R_3\cdot R_4\cdot
 R_3$ of $SU(3)_R\subset E_6$
 $$R=\pmatrix{1&0&0&0&0&0\cr
 0&1&1&1&0&0\cr
 0&0&0&-1&0&0\cr
 0&0&-1&0&0&0\cr
 0&0&1&1&1&0\cr
 0&0&1&1&0&1\cr}\ ,\eqno(21)$$
 which exchanges $Q_1\leftrightarrow Q_2$
 and leaves the SM charges unchanged (for the
 fields in the {\bf 27} irrep, $R$ transforms
 $d^c\leftrightarrow D^c$, $h'^-\leftrightarrow e$,
 $h'^0\leftrightarrow \nu$, and $\nu_4\leftrightarrow \nu_5$
 leaving the rest unchanged). Therefore,
 $g^{(1)}_2$ and $g^{(2)}_2$, as well as
 $g^{(1)}_3$ and $g^{(2)}_3$, define identical models
 related by $R$. There are
 two $Z_2$ and three $Z_3$ non-equivalent discrete gauge symmetries
 based on $E_6$ (see Table VI).
 Each one implies a low energy model with a
 definite pattern of fields and couplings in the superpotential. In
 Table VI we named $Z_2^a=g^{(1)}_2,\ Z_2^b=g^{(1)}_2g^{(2)}_2,\
 Z_3^a=g^{(1)}_3,\ Z_3^b=g^{(1)}_3g^{(2)}_3$ and
 $Z_3^c=g^{(1)}_3(g_3^{(2)})^{-1}$, where we identify the
 discrete symmetry group with the element that generates it.

The generic expression for a weight vector of $E_6$ whose VEV
 respects the SM symmetry is
 $$H_{n,m}=(n\! +\! m\ \ -\! m\ \ -\! n\ \ m\ \ -\! m\ \ n)\ ,\eqno(22)$$
 where $m$ and $n$ are integers.
 The $Q_1$ and $Q_2$ charges of such a weight of $E_6$
 are
 $$Q_1(H_{n,m})=n\eqno(23)$$
 and
 $$Q_2(H_{n,m})=m\ .\eqno(24)$$
 To reduce the rank of the gauge group from
 six to four one needs the VEVs of at least two flavors
 $H_{n,m}$ and $H_{n',m'}$ such that
 $\ {n m'} \ne {m n'}\ $.

The discrete symmetries $Z_2^a$ and $Z_3^a$, included in
 $U(1)_{Q_1}$, are left unbroken by Higgs $H_{n,m}$ satisfying
 $\ n=0\ mod\ 2\ $ and $\ n=0\ mod\ 3\ $, respectively.
 One may obtain a $Z_2^a$ model, for example, combining the
 VEVs of $H_{0,1}$, the $\nu_5$ flavour in the ${\bf 27}$
 or in the ${\bf 351}=(0\ 0\ 0\ 1\ 0\ 0)$ representations,
 and $H_{2,0}$ in the $\overline{\bf 351'}=(2\ 0\ 0\ 0\ 0\ 0)$
 or the ${\bf 2430}=(0\ 0\ 0\ 0\ 0\ 2)$. The symmetry
 $Z_3^a$ results from VEVs of $H_{1,0}$ and
 $H_{3,0}$, in the ${\bf 3003}=(3\ 0\ 0\ 0\ 0\ 0)$ or
 the $\overline{\bf 112320}=(1\ 1\ 0\ 0\ 1\ 0)$. In both cases
 these are the lowest dimensional representations required.
 Note that the same results would be obtained
 from the Weyl reflection $H_{n,m}\leftrightarrow H_{m,n}$ of
 VEVs in Eq.~(21) that, in particular,
 exchanges the $\nu_5$ and $\nu_4$ flavours in the ${\bf 27}$.
 Concerning the low energy implications,
 $Z_2^a$ and $Z_3^a$ are equivalent to the discrete symmetries in
 $SO(10)$ previously discussed (see Section 2), and they imply the
 same type of models.

The symmetry $Z^b_2$ in Table VI survives if all the GUT Higgs fields
 $H_{n,m}$ satisfy
 $$n+m=0\ mod\ 2\ .\eqno(25)$$
 This occurs, for example, by taking $H_{1,-1}$ in the
 ${\bf 78}=(0\ 0\ 0\ 0\ 0\ 1)$ or the ${\bf 650}=(1\ 0\ 0\ 0\ 1\ 0)$
 irreps of $E_6$,
 with $H_{1,1}$ in the
 $\overline{\bf 351}=(0\ 1\ 0\ 0\ 0\ 0)$ or the
 $\overline{\bf 351'}$, or with $H_{2,0}$ or
 $H_{0,2}$ also in the $\overline{\bf 351'}$ irrep.

 $E_6$ models with an unbroken $Z^b_2$ symmetry contain at low
 energies one extra pair of doublets $(h,h')$ and of
 singlets $(D,D^c)$ for each family of quarks and leptons; the
 neutrinos $(\nu_4, \nu_5)$ are not protected by $Z^b_2$ from heavy
 mass contributions (see Table VI). Although such a spectrum is
 compatible with perturbative unification (actually, it has been
 shown [12] that the gauge couplings unify at $\sim 10^{17} GeV$ with
 an electroweak mixing angle $\sin^2\theta_W=0.23$), these models predict
 unsuppressed proton decay mediated by the SUSY partners of
 $d^c$ and $D^c$, since all the trilinears in Eq.~(10) are
 allowed by $Z^b_2$.

The discrete symmetries $Z^b_3$ and $Z^c_3$ survive the VEVs
 of Higgs fields $H_{n,m}$ such that
 $$n+m=0\ mod\ 3\eqno(26)$$
 and
 $$n-m=0\ mod\ 3\ ,\eqno(27)$$
 respectively. The symmetry
 $Z^b_3$ may result, for example, using the flavours
 $H_{1,-1}$ (see above) and $H_{1,2}$ in the $\overline
 {\bf 5824}=(1\ 1\ 0\ 0\ 0\ 0)$ (which also contains $H_{1,-1}$)
 or $H_{3,0}$ while for $Z^c_3$ one can take $H_{1,1}$ and
 $H_{1,-2}$, both in the $\overline {\bf 1728}=(0\ 0\ 0\ 0\ 1\ 1)$
 irrep. Both discrete symmetries protect the
 quark singlets $(D,D^c)$ and the lepton/Higgs doublets $(h,h')$
 in the ${\bf 27}$ representation from acquiring heavy masses.
 $Z^b_3$ predicts, in addition, one pair of neutrinos
 $(\nu_4,\nu_5)$ per family. The $Z^b_3$ case seems unrealistic
 since the presence of trilinears $\ u^c d^c d^c\ $ and
 $\ q d^c l\ $ in $P$ would produce too rapid proton decay.
 In $Z^c_3$ models, although all the dangerous terms in Eq.~(10)
 are absent, there is also an unacceptable proton decay rate
 due to processes with exchange of {\it squarks} $D$ and $D^c$.
 A possible way to alleviate this problem could be to assume that
 one of the pairs $(\nu_4,\nu_5)$ remains light at
 $M_{GUT}\sim 10^{17}GeV$ and the scalar $\tilde \nu_5$ develops
 an intermediate VEV $\sim 10^{11}GeV$. Such a VEV would break
 $Z^c_3$ and make massive all the extra quarks and lepton doublets
 through terms of type $\ D D^c \nu_5\ $ and $\ h h' \nu_5\ $.
 The terms in Eq.~(10)
 would then be present but suppressed by powers of
 ${{<\nu_5>}\over M_{GUT}}$, and the required additional suppression
 seems reasonable.
 Note also the absence in $Z^c_3$ models of
 the dimension five operator $\ qqql\ $ [4],
 which in R-parity symmetric models generates a proton decay
 amplitude suppressed only by $M^{-1}_{GUT}$.

There are in $E_6$ three types of subgroups which may define
satisfactory models: $G_1=SO(10)\times U(1)$,
$G_2=SU(3)_C\times SU(3)_L\times SU(3)_R$,
and $G_3=SU(6)\times SU(2)$.
$SO(10)$ has been previously considered, and the projection matrix
$P_{G_1\subset E_6}$ is given at the begining of this section.

For $G_2$, the projection matrix is
$$P_{E_6\rightarrow G_2}=\pmatrix{
1&2&2&1&0&1\cr
0&0&1&1&1&1\cr
1&1&1&1&1&0\cr
0&-\! 1&-\! 1&-\! 1&-\! 1&0\cr
0&0&1&0&0&0\cr
0&0&-\! 1&-\! 1&0&0\cr}\ .\eqno(28)$$
$G_2$ is a rank-six group, and includes the two extra $U(1)$
factors of $E_6$. The analysis of Higgs flavours that leave a
discrete symmetry unbroken can be done straightforwardly by
projecting the results obtained for $E_6$. It turns out that
$$H_{n,m}=
(0\ \ 0)_C\ (0\ \ n\! +\! m)_L\ (-\! n\ \ n\! -\! m)_R\ .\eqno(29)$$
The standard matter parity $Z^a_2$, for example, will be
obtained combining the VEVs of $H_{0,1}$ in the
$(0\ 0)(0\ 1)(1\ 0)=({\bf 1},\overline {\bf 3},{\bf 3})$
and $H_{2,0}$ in the
$(0\ 0)(0\ 2)(1\ 2)=({\bf 1},\overline{\bf 6},\overline
{\bf 15})$.

The projection matrix to $SU(6)\times SU(2)$ is
$$P_{E_6\rightarrow G_3}=\pmatrix{
-\! 1&-\! 1&-\! 1&-\! 1&0&-\! 1\cr
0&1&1&1&0&1\cr
1&1&1&0&0&0\cr
0&0&0&1&1&0\cr
0&0&1&0&0&1\cr
0&0&-\! 1&-\! 1&0&0\cr}\ .\eqno(30)$$
The projected Higgs $H_{n,m}$ has the form
$$H_{n,m}=(-\! n\! -\! m\ \ 0\ \ 0\ \ 0\ \ 0)(n\! -\! m)\ .\eqno(31)$$
To obtain a low energy limit with, for instance,
$Z^a_2$ discrete
symmetry one may use $H_{0,1}$ in the $(0\ 0\ 0\ 0\ 1)(1)
=(\overline{\bf 6},{\bf 2})$ with $H_{2,0}$ in the
$(0\ 0\ 0\ 0\ 2)(2)=(\overline{\bf 21},{\bf 3})$.
The projection from $SU(6)$ to $SU(5)$ is done by droping
the first component of the weight vector.

\medskip
\leftline {\bf 3. Summary and conclusions}
\medskip

We have analyzed the appearance of discrete symmetries
of the superpotential as remnants of gauge symmetries in
models based on $E_6$ and $SO(10)$. These models include
{\it flipped} and {\it unflipped} $SU(5)$, $SU(4)\!\times\!
SU(2)_L\!\times\! SU(2)_R$, $SU(3)_C\!\times\! SU(3)_L\!\times\!
SU(3)_R$, and $SU(6)\!\times\! SU(2)$. We have used the
Dynking labelling to identify the GUT Higgs leading to each
discrete gauge symmetry and have established the matter content
and couplings of the different low energy models. These type of
discrete symmetries are not anomalous respect to gravitational
effects, and in SUSY models could work as matter parities.

Our results are the following. In models based on $SO(10)$ there is
only one generic $Z_N$. The case $N=2$ corresponds to the usual
R-parity of SUSY scenarios, with minimal matter content, an stable
LSP, and absence of B and L violating terms in
the superpotential. For $N\ge 3$ the trilinears in $P$ involving
quarks and leptons are the same, but the presence of right handed
neutrinos may introduce LSP decay and L violation. All $Z_N$
symmetries would be broken if one uses as a Higgs the singlet
in the ${\bf 16}$ of $SO(10)$, which is the smallest representation
that reduces the rank of $SO(10)$ while preserving the SM symmetry.
The smallest representation containing a GUT Higgs whose
VEV breaks the extra $U(1)$ and safes a $Z_N$ symmetry is the
$(0\ 0\ 0\ 0\ N)$, whose dimension is given in Eq.~(8).

In models based on $E_6$ the discrete symmetry is a $Z_N\times
Z_M$. In addition to the $Z_N$ models of $SO(10)$, we
find here another $Z_2$ and two more $Z_3$ nonequivalent cases.
All these symmetries imply light nonstandard lepton/Higgs
doublets and down type quarks, and it is difficult (although in
one of the cases it seems possible) to accommodate a long enough
proton lifetime. In general, high dimensional representations
are required to obtain these symmetries; using as Higgs the
flavors $\nu_4$ and $\nu_5$ in the ${\bf 27}$ no gauge discrete
symmetry survives.

Obviously, there are other consistent choices of the GUT group.
One may consider large $SU(n)$ groups, but in general they
require complicated choices of representations
to cancel anomalies. One may as well consider non simple
groups, but then the perturbative unification of the gauge
couplings would be purely accidental. It is also possible
to consider the exceptional groups
$E_7$ and $E_8$ (with only real representations), $SO(32)$
(which contains the $SU(15)$ model [13]), or family models
based on $SO(4n+2)$, but all of them
predict the presence at low energies
of mirror partners instead of the observed three chiral families.
If one thinks of a SUSY extension of the SM embedded in
a chiral GUT, then the only models based on discrete
gauge symmetries (non anomalous
respect gravitational effects) are the ones found here.

\medskip
\leftline {\bf Acknowledgments}
\medskip

The authors are indebted to P. Griffin and R.T. Sharp for helpful
discussions and a reading of the manuscript.
This work has been supported by the Natural Sciences and Engineering
Research Council of Canada (M.~De M.) and the
Ministerio de Educaci\'on y Ciencia of Spain (M.~M.).

\medskip
\leftline {\bf References}
\medskip

\noindent [1] G. Farrar and P. Fayet, {\it Phys. Lett.\/}
 {\bf B76\/}, 575 (1978); S. Dimopoulos and H. Georgi,
 {\it Nucl. Phys.\/} {\bf B193\/}, 150 (1981); S. Weinberg,
 {\it Phys. Rev.\/} {\bf D26\/}, 287 (1982); for a review
 on Supersymmetry, see H.P. Nilles, {\it Phys. Rep.}
 {\bf 110} 1 (1984).

\noindent [2] M. C. Bento, L. Hall and G. G. Ross,
 {\it Nucl. Phys.\/} {\bf B292\/}, 400 (1987); H. Dreiner
 and G.G. Ross, {\it Nucl. Phys.} {\bf B365\/}, 597 (1991).

\noindent [3] L. E. Ib\'a\~nez and G. G. Ross,
 {\it Phys. Lett.\/} {\bf B260\/}, 291 (1991);
 {\it Nucl. Phys.\/} {\bf B368\/}, 3 (1992).

\noindent [4] For a review on GUTs, see
 G.G. Ross, {\it Grand Unified Theories}
 (Benjamin-Cummings, Redwood City, CA, 1985).

\noindent [5] L. Krauss and F. Wilczek, {\it Phys. Rev. Lett.\/}
 {\bf 62\/}, 1221 (1989).

\noindent [6] S. P. Martin, {\it Phys. Rev.\/} {\bf D46\/},
 R2769 (1992).

\noindent [7] E. B. Dynkin, {\it Am. Math. Soc. Transl.\/},
 Ser. 2, {\bf 6\/}, 111; 245 (1957);
 R. Cahn, {\it Semi Simple Lie Algebras and Their
 Representations\/} (Benjamin-Cummings, New York, 1984);
 S. Kass, R. V. Moody, J. Patera and R. Slansky,
 {\it Affine Lie Algebras, Weight Multiplicities and Branching
 Rules\/}, Vol 1 (University of California Press, 1990).

\noindent [8] R. Slansky, {\it Phys. Rep.\/} {\bf 79\/}, 1 (1981).

\noindent [9] I. Antoniadis, J. Ellis, J.S. Hagelin,
and D.V. Nanopoulos, {\it Phys. Lett.\/} {\bf B194\/}, 231 (1987).

\noindent [10] M. Masip, {\it Phys. Rev.\/} {\bf D47\/},
 3071 (1993);
M. Masip and Y. Wang, {\it Phys. Rev.\/} {\bf D48\/}, 1555 (1993).

\noindent [11] W. McKay, J. Patera and D. Sankoff, in
 {\it Computers in Non Associative Rings and Algebras\/},
 eds. R. Beck and B. Kolman, Academic Press, New York, 235 (1977).

\noindent [12] F. del Aguila, G.D. Coughlan, and M. Masip,
 {\it Nucl. Phys.\/} {\bf B351\/}, 90 (1991).

\noindent [13] S.L. Adler, {\it Phys. Lett.\/} {\bf B225\/},
 143 (1989).

\vfil\eject

\leftskip=1.8pc
\rightskip=1.pc
\baselineskip=24truept

\noindent {\bf Table I.} ${\bf 16}=(0\ 0\ 0\ 0\ 1)$ irrep of $SO(10)$
and nonstandard $E$ charge.
$$
\baselineskip=10truept
\vbox{
\tabskip=.6truecm
\halign{#&\hfil#\hfil&\hfil#\hfil&#&\hfil#\hfil \cr
\noalign{\hrule}
\omit&\omit&\omit&\omit&\omit\cr
\omit&\omit&\omit&\omit&\omit\cr
&$E$&\ &&$E$\cr
\omit&\omit&\omit&\omit&\omit\cr
\omit&\omit&\omit&\omit&\omit\cr
\noalign{\hrule}
\omit&\omit&\omit&\omit&\omit\cr
\omit&\omit&\omit&\omit&\omit\cr
$u:(0\ 0\ 0\ 0\ 1)$&$-1$&&
$u^c:(1\ 0\ -\! 1\ 0\ 1)$&$-1$\cr
\omit&\omit&\omit&\omit&\omit\cr
\omit&\omit&\omit&\omit&\omit\cr
$d^c:( 0\ 0\ 1\ 0\ -\! 1)$&$3$&&
$N:(-\! 1\ 1\ -\! 1\ 0\ 1)$&$-5$\cr
\omit&\omit&\omit&\omit&\omit\cr
\omit&\omit&\omit&\omit&\omit\cr
$u^c:(0\ 1\ -\! 1\ 1\ 0)$&$-1$&&
$e:(1\ 0\ 0\ 0\ -\! 1)$&$3$\cr
\omit&\omit&\omit&\omit&\omit\cr
\omit&\omit&\omit&\omit&\omit\cr
$\nu:(1\ -\! 1\ 0\ 1\ 0)$&$3$&&
$u:(0\ -\! 1\ 0\ 0\ 1)$&$-1$\cr
\omit&\omit&\omit&\omit&\omit\cr
\omit&\omit&\omit&\omit&\omit\cr
$d:(0\ 1\ 0\ -\! 1\ 0)$&$-1$&&
$d:(-\! 1\ 1\ 0\ 0\ -\! 1)$&$-1$\cr
\omit&\omit&\omit&\omit&\omit\cr
\omit&\omit&\omit&\omit&\omit\cr
$u:(-\! 1\ 0\ 0\ 1\ 0)$&$-1$&&
$d^c:(0\ -\! 1\ 1\ 0\ -\! 1)$&$-3$\cr
\omit&\omit&\omit&\omit&\omit\cr
\omit&\omit&\omit&\omit&\omit\cr
$d^c:(1\ -\! 1\ 1\ -\! 1\ 0)$&$3$&&
$u^c:(0\ 0\ -\! 1\ 1\ 0)$&$-1$\cr
\omit&\omit&\omit&\omit&\omit\cr
\omit&\omit&\omit&\omit&\omit\cr
$e^c:(-\! 1\ 0\ 1\ -\! 1\ 0)$&$-1$&&
$d:(0\ 0\ 0\ -\! 1\ 0)$&$-1$\cr
\omit&\omit&\omit&\omit&\omit\cr
\omit&\omit&\omit&\omit&\omit\cr
\noalign{\hrule}
}}$$
\vskip 2truecm

\noindent {\bf Table II.} ${\bf 10}=(1\ 0\ 0\ 0\ 0)$ irrep of $SO(10)$
and nonstandard $E$ charge.
$$
\baselineskip=10truept
\vbox{
\tabskip=.6truecm
\halign{#&\hfil#\hfil&\hfil#\hfil&#&\hfil#\hfil \cr
\noalign{\hrule}
\omit&\omit&\omit&\omit&\omit\cr
\omit&\omit&\omit&\omit&\omit\cr
&$E$&\ &&$E$\cr
\omit&\omit&\omit&\omit&\omit\cr
\omit&\omit&\omit&\omit&\omit\cr
\noalign{\hrule}
\omit&\omit&\omit&\omit&\omit\cr
\omit&\omit&\omit&\omit&\omit\cr
$D:(1\ 0\ 0\ 0\ 0)$&$2$&&
$D:(0\ 0\ 0\ 1\ -\! 1)$&$2$\cr
\omit&\omit&\omit&\omit&\omit\cr
\omit&\omit&\omit&\omit&\omit\cr
$D^c:(-\! 1\ 1\ 0\ 0\ 0)$&$-2$&&
$h^0:(0\ 0\ 1\ -\! 1\ -\! 1)$&$2$\cr
\omit&\omit&\omit&\omit&\omit\cr
\omit&\omit&\omit&\omit&\omit\cr
$h^+:(0\ -\! 1\ 1\ 0\ 0)$&$2$&&
$h'^-:(0\ 1\ -\! 1\ 0\ 0)$&$-2$\cr
\omit&\omit&\omit&\omit&\omit\cr
\omit&\omit&\omit&\omit&\omit\cr
$h'^0:(0\ 0\ -\! 1\ 1\ 1)$&$-2$&&
$D:(1\ -\! 1\ 0\ 0\ 0)$&$2$\cr
\omit&\omit&\omit&\omit&\omit\cr
\omit&\omit&\omit&\omit&\omit\cr
$D^c:(0\ 0\ 0\ -\! 1\ 1)$&$-2$&&
$D^c:(-\! 1\ 0\ 0\ 0\ 0)$&$-2$\cr
\omit&\omit&\omit&\omit&\omit\cr
\omit&\omit&\omit&\omit&\omit\cr
\noalign{\hrule}
}}$$

\vfil\eject

\noindent {\bf Table III.}
Generic $Z_N$ discrete symmetry
of models based on $SO(10)$ ($\alpha^N=1$)
and the particular cases with $N=2$ and $N=3$ ($\sigma^3=1$).
$$
\baselineskip=10truept
\vbox{
\tabskip=.6truecm
\halign{\hfil#\hfil&\hfil#\hfil&\hfil#\hfil&\hfil#\hfil
&\hfil#\hfil &\hfil#\hfil &\hfil#\hfil &\hfil#\hfil &\hfil#\hfil
&\hfil#\hfil &\hfil#\hfil \cr
\noalign{\hrule}
\omit&\omit&\omit&\omit&\omit&\omit&\omit&\omit&\omit&\omit&\omit\cr
\omit&\omit&\omit&\omit&\omit&\omit&\omit&\omit&\omit&\omit&\omit\cr
&$q$&$u^c$&$d^c$&$l$&$e^c$&$N$&$h$&$h'$&$D^c$&$D$\cr
\omit&\omit&\omit&\omit&\omit&\omit&\omit&\omit&\omit&\omit&\omit\cr
\omit&\omit&\omit&\omit&\omit&\omit&\omit&\omit&\omit&\omit&\omit\cr
\noalign{\hrule}
\omit&\omit&\omit&\omit&\omit&\omit&\omit&\omit&\omit&\omit&\omit\cr
\omit&\omit&\omit&\omit&\omit&\omit&\omit&\omit&\omit&\omit&\omit\cr
$Z_N$&$1$&$\alpha$&$\alpha^{-1}$&$1$&$\alpha^{-1}$&$\alpha$&
$\alpha^{-1}$&$\alpha$&$1$&$1$\cr
\omit&\omit&\omit&\omit&\omit&\omit&\omit&\omit&\omit&\omit&\omit\cr
\omit&\omit&\omit&\omit&\omit&\omit&\omit&\omit&\omit&\omit&\omit\cr
$Z_2$&$1$&$-1$&$-1$&$1$&$-1$&$-1$&$-1$&$-1$&$1$&$1$\cr
\omit&\omit&\omit&\omit&\omit&\omit&\omit&\omit&\omit&\omit&\omit\cr
\omit&\omit&\omit&\omit&\omit&\omit&\omit&\omit&\omit&\omit&\omit\cr
$Z_3$&$1$&$\sigma$&$\sigma^2$&$1$&$\sigma^2$&$\sigma$&
$\sigma^2$&$\sigma$&$1$&$1$\cr
\omit&\omit&\omit&\omit&\omit&\omit&\omit&\omit&\omit&\omit&\omit\cr
\omit&\omit&\omit&\omit&\omit&\omit&\omit&\omit&\omit&\omit&\omit\cr
\noalign{\hrule}
}}$$
\vskip 2truecm

\noindent {\bf Table IV.}
List of $SO(10)$ irreps which contain a Higgs $H_n$ preserving a
a $Z_n$ discrete symmetry for $n=1,2,3$.
$$
\baselineskip=10truept
\vbox{
\tabskip=.6truecm
\halign{\hfil#\hfil&\hfil#\hfil \cr
\noalign{\hrule}
\omit&\omit\cr
\omit&\omit\cr
$H_1=(-\! 1\ 1\ -\! 1\ 0\ 1)$&${\bf 16}=(0\ 0\ 0\ 0\ 1)$\cr
\omit&\omit\cr
\omit&\omit\cr
&${\bf 144}=(1\ 0\ 0\ 1\ 0)$\cr
\omit&\omit\cr
\omit&\omit\cr
&${\bf 560}=(0\ 1\ 0\ 0\ 1)$\cr
\omit&\omit\cr
\omit&\omit\cr
\noalign{\hrule}
\omit&\omit\cr
\omit&\omit\cr
$H_2=(-\! 2\ 2\ -\! 2\ 0\ 2)$&${\bf 126}=(0\ 0\ 0\ 0\ 2)$\cr
\omit&\omit\cr
\omit&\omit\cr
&${\bf 1728}=(1\ 0\ 0\ 1\ 1)$\cr
\omit&\omit\cr
\omit&\omit\cr
&${\bf 2970}=(0\ 1\ 1\ 0\ 0)$\cr
\omit&\omit\cr
\omit&\omit\cr
\noalign{\hrule}
\omit&\omit\cr
\omit&\omit\cr
$H_3=(-\! 3\ 3\ -\! 3\ 0\ 3)$&
$\overline{\bf 672}=(0\ 0\ 0\ 0\ 3)$\cr
\omit&\omit\cr
\omit&\omit\cr
&$\overline{\bf 11088}=(1\ 0\ 0\ 2\ 1)$\cr
\omit&\omit\cr
\omit&\omit\cr
&${\bf 49280}=(2\ 0\ 0\ 2\ 1)$\cr
\omit&\omit\cr
\omit&\omit\cr
\noalign{\hrule}
}}$$

\vfill\eject

\noindent {\bf Table V.} ${\bf 27}=(1\ 0\ 0\ 0\ 0\ 0)$ irrep of $E_6$
and nonstandard $E$ and $F$ charges.
$$
\baselineskip=10truept
\vbox{
\tabskip=.6truecm
\halign{#&\hfil#\hfil&\hfil#\hfil&\hfil#\hfil
&#&\hfil#\hfil&\hfil#\hfil \cr
\noalign{\hrule}
\omit&\omit&\omit&\omit&\omit&\omit&\omit\cr
\omit&\omit&\omit&\omit&\omit&\omit&\omit\cr
&$E$&$F$&\ &&$E$&$F$\cr
\omit&\omit&\omit&\omit&\omit&\omit&\omit\cr
\omit&\omit&\omit&\omit&\omit&\omit&\omit\cr
\noalign{\hrule}
\omit&\omit&\omit&\omit&\omit&\omit&\omit\cr
\omit&\omit&\omit&\omit&\omit&\omit&\omit\cr
$u:(1\ 0\ 0\ 0\ 0\ 0)$&$-1$&$1$&&
$D:(-\! 1\ 0\ 0\ 0\ 1\ 0)$&$2$&$-2$\cr
\omit&\omit&\omit&\omit&\omit&\omit&\omit\cr
\omit&\omit&\omit&\omit&\omit&\omit&\omit\cr
$D:( -\! 1\ 1\ 0\ 0\ 0\ 0)$&$2$&$-2$&&
$e:(-\! 1\ 0\ 0\ 1\ -\!\! 1\ 0)$&$3$&$1$\cr
\omit&\omit&\omit&\omit&\omit&\omit&\omit\cr
\omit&\omit&\omit&\omit&\omit&\omit&\omit\cr
$d^c:(0\ -\!\! 1\ 1\ 0\ 0\ 0)$&$3$&$1$&&
$e^c:\ (1\ -\!\! 1\ 1\ -\!\! 1\ 0\ 0)$&$-1$&$1$\cr
\omit&\omit&\omit&\omit&\omit&\omit&\omit\cr
\omit&\omit&\omit&\omit&\omit&\omit&\omit\cr
$u^c:\ (0\ 0\ -\!\! 1\ 1\ 0\ 1)$&$-1$&$1$&&
$h^0:(-\! 1\ 0\ 1\ -\! 1\ 0\ 0)$&$2$&$-2$\cr
\omit&\omit&\omit&\omit&\omit&\omit&\omit\cr
\omit&\omit&\omit&\omit&\omit&\omit&\omit\cr
$D^c:(0\ 0\ 0\ -\! 1\ 1\ 1)$&$-2$&$-2$&&
$\nu_4:(1\ 0\ -\! 1\ 0\ 0\ 1)$&$-5$&$1$\cr
\omit&\omit&\omit&\omit&\omit&\omit&\omit\cr
\omit&\omit&\omit&\omit&\omit&\omit&\omit\cr
$\nu:\ (0\ 0\ 0\ 1\ 0\ -\!\! 1)$&$3$&$1$&&
$h'^-:(-\! 1\ 1\ -\! 1\ 0\ 0\ 1)$&$-2$&$-2$\cr
\omit&\omit&\omit&\omit&\omit&\omit&\omit\cr
\omit&\omit&\omit&\omit&\omit&\omit&\omit\cr
$d:\ (0\ 0\ 0\ 0\ -\!\! 1\ 1)$&$-2$&$1$&&
$u:\ (1\ 0\ 0\ 0\ 0\ -\!\! 1)$&$-1$&$1$\cr
\omit&\omit&\omit&\omit&\omit&\omit&\omit\cr
\omit&\omit&\omit&\omit&\omit&\omit&\omit\cr
$h^+:(0\ 0\ 1\ -\! 1\ 1\ -\! 1)$&$2$&$-2$&&
$d:(0\ -\! 1\ 0\ 0\ 0\ 1)$&$-1$&$1$\cr
\omit&\omit&\omit&\omit&\omit&\omit&\omit\cr
\omit&\omit&\omit&\omit&\omit&\omit&\omit\cr
$d^c:(0\ 0\ 1\ 0\ -\! 1\ -\! 1)$&$3$&$1$&&
$D:(-\! 1\ 1\ 0\ 0\ 0\ -\! 1)$&$2$&$-2$\cr
\omit&\omit&\omit&\omit&\omit&\omit&\omit\cr
\omit&\omit&\omit&\omit&\omit&\omit&\omit\cr
$h'^0:(0\ 1\ -\! 1\ 0\ 1\ 0)$&$-2$&$-2$&&
$d^c:(0\ -\! 1\ 1\ 0\ 0\ -\! 1)$&$3$&$1$\cr
\omit&\omit&\omit&\omit&\omit&\omit&\omit\cr
\omit&\omit&\omit&\omit&\omit&\omit&\omit\cr
$u^c:(0\ 1\ -\! 1\ 1\ -\! 1\ 0)$&$-1$&$1$&&
$u^c:(0\ 0\ -\! 1\ 1\ 0\ 0)$&$-1$&$1$\cr
\omit&\omit&\omit&\omit&\omit&\omit&\omit\cr
\omit&\omit&\omit&\omit&\omit&\omit&\omit\cr
$u:\ (1\ -\! 1\ 0\ 0\ 1\ 0)$&$-1$&$1$&&
$D^c:(0\ 0\ 0\ -\! 1\ 1\ 0)$&$-2$&$-2$\cr
\omit&\omit&\omit&\omit&\omit&\omit&\omit\cr
\omit&\omit&\omit&\omit&\omit&\omit&\omit\cr
$\nu_5:(1\ -\!\! 1\ 0\ 1\ -\!\! 1\ 0)$&$0$&$4$&&
$d:(0\ 0\ 0\ 0\ -\! 1\ 0)$&$-1$&$1$\cr
\omit&\omit&\omit&\omit&\omit&\omit&\omit\cr
\omit&\omit&\omit&\omit&\omit&\omit&\omit\cr
$D^c:(0\ 1\ 0\ -\! 1\ 0\ 0)$&$-2$&$-2$&&&&\cr
\omit&\omit&\omit&\omit&\omit&\omit&\omit\cr
\omit&\omit&\omit&\omit&\omit&\omit&\omit\cr
\noalign{\hrule}
}}$$

\vfil\eject

\noindent {\bf Table VI.} Generators $g^{(1)}_N$ and $g^{(2)}_M$ of
the
$Z_N\times Z_M$ discrete symmetry ($\alpha^N=\beta^M=1$)
of models based on $E_6$, and
nonequivalent $Z_2$ and $Z_3$ ($\sigma^3=1$) particular cases.
$$
\baselineskip=10truept
\vbox{
\tabskip=.6truecm
\halign{\hfil#\hfil&\hfil#\hfil&\hfil#\hfil&\hfil#\hfil
&\hfil#\hfil &\hfil#\hfil &\hfil#\hfil &\hfil#\hfil &\hfil#\hfil
&\hfil#\hfil &\hfil#\hfil &\hfil#\hfil \cr
\noalign{\hrule}
\omit&
\omit&\omit&\omit&\omit&\omit&\omit&\omit&\omit&\omit&\omit&\omit\cr
\omit&
\omit&\omit&\omit&\omit&\omit&\omit&\omit&\omit&\omit&\omit&\omit\cr
&$q$&$u^c$&$d^c$&$l$&$e^c$&$h$&$h'$&$D^c$&$D$&$\nu_4$&$\nu_5$\cr
\omit&
\omit&\omit&\omit&\omit&\omit&\omit&\omit&\omit&\omit&\omit&\omit\cr
\omit&
\omit&\omit&\omit&\omit&\omit&\omit&\omit&\omit&\omit&\omit&\omit\cr
\noalign{\hrule}
\omit&
\omit&\omit&\omit&\omit&\omit&\omit&\omit&\omit&\omit&\omit&\omit\cr
\omit&
\omit&\omit&\omit&\omit&\omit&\omit&\omit&\omit&\omit&\omit&\omit\cr
$g^1_N$&$1$&$\alpha$&$\alpha^{-1}$&$1$&$\alpha^{-1}$
&$\alpha^{-1}$&$\alpha$&$1$&$1$&$\alpha$&$1$\cr
\omit&
\omit&\omit&\omit&\omit&\omit&\omit&\omit&\omit&\omit&\omit&\omit\cr
\omit&
\omit&\omit&\omit&\omit&\omit&\omit&\omit&\omit&\omit&\omit&\omit\cr
$g^2_M$&$1$&$\beta$&$1$&$\beta$&$\beta^{-1}$&$\beta^{-1}$&$1$&
$\beta^{-1}$&$1$&$1$&$\beta$\cr
\omit&
\omit&\omit&\omit&\omit&\omit&\omit&\omit&\omit&\omit&\omit&\omit\cr
\omit&
\omit&\omit&\omit&\omit&\omit&\omit&\omit&\omit&\omit&\omit&\omit\cr
$Z^a_2$&$1$&$-1$&$-1$&$1$&$-1$&$-1$&$-1$&$1$&$1$&$-1$&$1$\cr
\omit&
\omit&\omit&\omit&\omit&\omit&\omit&\omit&\omit&\omit&\omit&\omit\cr
\omit&
\omit&\omit&\omit&\omit&\omit&\omit&\omit&\omit&\omit&\omit&\omit\cr
$Z^b_2$&$1$&$1$&$-1$&$-1$&$1$&$1$&$-1$&$-1$&$1$&$-1$&$-1$\cr
\omit&
\omit&\omit&\omit&\omit&\omit&\omit&\omit&\omit&\omit&\omit&\omit\cr
\omit&
\omit&\omit&\omit&\omit&\omit&\omit&\omit&\omit&\omit&\omit&\omit\cr
$Z^a_3$&$1$&$\sigma$&$\sigma^2$&$1$&$\sigma^2$&
$\sigma^2$&$\sigma$&$1$&$1$&$\sigma$&$1$\cr
\omit&
\omit&\omit&\omit&\omit&\omit&\omit&\omit&\omit&\omit&\omit&\omit\cr
\omit&
\omit&\omit&\omit&\omit&\omit&\omit&\omit&\omit&\omit&\omit&\omit\cr
$Z^b_3$&$1$&$\sigma^2$&$\sigma^2$&$\sigma$&$\sigma$&
$\sigma$&$\sigma$&$\sigma^2$&$1$&$\sigma$&$\sigma$\cr
\omit&
\omit&\omit&\omit&\omit&\omit&\omit&\omit&\omit&\omit&\omit&\omit\cr
\omit&
\omit&\omit&\omit&\omit&\omit&\omit&\omit&\omit&\omit&\omit&\omit\cr
$Z^c_3$&$1$&$1$&$\sigma^2$&$\sigma^2$&$1$&
$1$&$\sigma$&$\sigma$&$1$&$\sigma$&$\sigma^2$\cr
\omit&
\omit&\omit&\omit&\omit&\omit&\omit&\omit&\omit&\omit&\omit&\omit\cr
\omit&
\omit&\omit&\omit&\omit&\omit&\omit&\omit&\omit&\omit&\omit&\omit\cr
\noalign{\hrule}
}}$$
\vfil\eject\end